# Karman constant and accurate mean flow prediction in a turbulent pipe

Authors:    Zhen-Su She*[1], Xi Chen[1], You Wu[1], Fazle Hussain[1,2]

*Affiliations:*

[1]*State Key Laboratory for Turbulence and Complex Systems and Department of Mechanics, College of Engineering, Peking University, Beijing, 100871, China*

[2]*Department of Mechanical Engineering, University of Houston, Houston, TX, 77204-4006, USA*

* *Email address:    she@pku.edu.cn*

1 **Summary**

2 The Karman constant $\kappa$ - widely used in atmospheric science and engineering
3 turbulence modelling, and proposed by Prandtl[1] in 1925 and von Karman[2] in 1930 to
4 describe the mean velocity of a turbulent wall-bounded flow[3] - leads to a logarithmic
5 profile in an overlap region near the wall. For over eighty years, its value was believed
6 to be ~0.41. But more recently, many argue that it is not a constant, because of
7 measured variations in different flows and at different Reynolds numbers (*Re*)[4-6]. Here,
8 a multi-layer analytic theory[7] is shown to lead to a re-interpretation of $\kappa$ as a global
9 constant for both the overlap region and outer flow, and to yield a new method for its
10 measurement. The newly determined value is 0.45 for both channel and pipe. It is
11 shown that this new $\kappa$, together with other wall constants, yields a 99% accuracy in the
12 prediction of mean velocity data at *all points* in high *Re* (up to 40 million) pipe flow.
13 The theory also describes finite *Re* effect, and discovers a transition at the friction *Re*
14 (i.e. $\mathrm{Re}_\tau$) = 5000. An accurate model for the prediction of turbulent transport in
15 canonical pipe and channel flows is achieved here, and we propose the model to be
16 valid for a wide class of turbulent flows.

**Text**

Accurate prediction of turbulent flow remains a top challenge in classical physics. Even for simple wall-bounded flows such as pipe and channel, analytic prediction of mean velocities at very high *Re* relies on empirical functions[8] with compromised accuracy, and continues to receive vivid attention with great experimental efforts[9-14]. Deriving a mean-field prediction based on first principles is highly desired, as it would reveal new statistical symmetries governing turbulent fluctuations and allow a deeper understanding of important constants describing the flow.

Prandtl[1] in 1925, and von Karman[2] in 1930, independently suggested the concept of *mixing length* - analogous to the mean free path for molecular collisions - to model turbulent transport. A simple model of the mixing length with linear dependence on the distance from the wall yields a logarithmic mean velocity profile (MVP), where a proportionality constant - the Karman constant $\kappa$ - was introduced. However, this empirical model has led to controversies: Barenblatt and co-authors[15-16] have claimed that power-law is a better description. Indeed, theoretical interpretation of the log-law is very unsatisfactory. For example, very little is understood about $\kappa$ and the other log-law constant $B$; their variability with different flows and different *Re* remains elusive[4-6]. While the log-law vs. power law debate is still vivid[17-18], we present a coveted theory which re-interprets $\kappa$ with a new method for its measurement. Not only a more accurate description of the entire MVP over a wide range of *Re* is obtained, the theory also predicts, for the first time, a critical *Re* in Princeton pipe experiment, and derives quantities of engineering interest. The theory presented here provides a general guide

40  for accurate mean-field prediction from post-analysis of massive simulation data for a
41  wide class of turbulent flows.
42     The mean velocity in a turbulent pipe flow has resisted analytic attack since the turn
43  of the last century[9]. We overcome this bottleneck by developing a symmetry analysis on
44  the mean momentum equation (MME), utilizing empirical knowledge about the
45  multiple layers. For a channel or a pipe, the MME has an exact first-integral[19], written
46  in wall units as

$$S^+ + W^+ = 1 - y^+ / \text{Re}_\tau = r, \tag{1}$$

48  where $S^+ = dU^+/dy^+$ is the mean shear and $W^+ = -\langle u'v' \rangle^+$ is the Reynolds stress,
49  and $y^+$ the distance to the wall, r the distance to the center, $\text{Re}_\tau$ the friction Reynolds
50  number, and + denotes normalization using wall units. Our symmetry analysis involves
51  a length determined by the two primary variables, $W^+$ and $S^+$, which is, by
52  dimensional argument, $\ell_M^+ = \sqrt{W^+}/S^+$. This length is regarded[7] as an "order
53  parameter" used to represent effects of fluctuations in statistical mean-field theory[20],
54  and happens to coincide with the mixing length[1,2] for describing the vertical mean
55  momentum transport. The concept of length order parameter is more general, however,
56  and is applicable to analyze other flows.
57     The symmetry analysis of the multi-layer structure is accomplished by a Lie-group
58  theory[7] which derives the entire profile of the mixing length in two steps: First, identify
59  local invariant solutions for each of the layers: well known as sublayer, buffer layer,
60  bulk zone, and a new central zone called 'core'. Each of the layers is postulated to
61  possess a local scaling for the mixing length or its gradient under the dilation

62   transformation. Then, the theory[7] postulates a transition ansatz and derives a combined,

63   analytic expression for the entire profile of the mixing length:

$$\ell_M^+ = \rho(\frac{y^+}{y_{sub}^+})^{3/2}\left(1+(\frac{y^+}{y_{sub}^+})^{p_I}\right)^{1/(2p_I)}\left(1+(\frac{y^+}{y_{buf}^+})^{p_{II}}\right)^{-1/p_{II}}\frac{1-r^m}{m(1-r)Z_{core}}\left(1+(\frac{r}{r_{core}})^{-p_{IV}}\right)^{1/(2p_{IV})}. \quad (2)$$

65   This determines the mean shear profile[19]

$$S^+ = (-1+\sqrt{4r\ell_M^{+2}+1})/(2\ell_M^{+2}), \quad (3)$$

67   hence the mean velocity profile $U^+ = \int S^+ dy^+$, the average velocities

68   ($\overline{U^+}^{Pipe} = 2\int U^+ r dr$ and $\overline{U^+}^{CH} = \int U^+ dr$), and the friction factor $C_f = 8/\overline{U^+}^2$. While

69   previous work reported heuristic and systematic derivation[7] of Eq.(2), the main task

70   here is to determine the parameters with a systematic method, and to demonstrate that

71   the Karman constant is indeed a constant, and that there is a transition at $\text{Re}_\tau = 5000$.

72       Two sets of parameters appear in Eq.(2). The *first* are four coefficients:

73   $\rho$, $y_{sub}^+$, $y_{buf}^+$, $r_{core}$; the *second* are called transition sharpnesses: $p_I$, $p_{II}$, $p_{IV}$, which

74   describe the transition between sublayer and buffer, buffer and log region, and bulk and

75   core. The second set, less sensitive to the predictions, are chosen to be integers

76   ($p_I = p_{II} = 4$, $p_{IV} = 2$). We focus below on the determination of the four coefficients.

77       First, the inner solution near the wall determines $\rho$ and $y_{sub}^+$. From Eq.(2), we

78   find that for $y^+ \ll y_{sub}^+$, $\ell_M^+ \approx \rho(y^+/y_{sub}^+)^{3/2}$. If the local scaling exists,

79   $\ell_M^+/y^{+3/2} \approx \rho(y_{sub}^+)^{-3/2}$ is a constant near the wall, determined by $\rho$ and $y_{sub}^+$. Fig.1

80   shows that well resolved direct numerical simulation (DNS) data for channel flow

81   indeed confirms the existence of the constant, with a value of 0.0315. In order to

82   determine $y_{sub}^+$, note that it is the location where the local scaling is in the middle of 3/2

83   (sublayer scaling) and 2 (buffer scaling); thus, $d\ln \ell_M^+ / d\ln y^+ |_{y_{sub}^+} = (1+3/2)/2 = 7/4$.

84   Furthermore, we assume, with accurate empirical support from DNS data, that this

85   transition point coincides with the maximum of $\Gamma = S^+ y^+$, the diagnostic function used

86   to quantify near-wall statistics[4]. With the assumption $d\Gamma(y_{sub}^+)/dy^+ = 0$, we

87   differentiate Eq.(1) at $y^+ = y_{sub}^+$ to derive the following appealing results:

88   $S^+(y_{sub}^+) = 3/5$, $W^+(y_{sub}^+) = 2/5$, $\ell_M^+(y_{sub}^+) = \sqrt{10}/3$. This yields the prediction:

89   $\rho_0 \approx 2^{3/8}\sqrt{5}/3 \approx 0.967$; then, using the above empirical near-wall measurement, we

90   obtain $y_{sub}^+ \approx 9.80$.

91       Next, Eq.(2) predicts an analytic scaling function for the mean defect velocity (i.e.

92   difference from the centreline velocity $U_c^+$) (see Fig.2). Fig.2a confirms this linear

93   relation for channel (DNS) and pipe (Princeton data) over a wide range of *Re* (a

94   complete collapse); then, $\kappa$ is derived from the slope with high accuracy. Fig.2b

95   shows a procedure for the determination of $r_{core}$. A conservative estimate of the

96   Karman constant is $\kappa \approx 0.45 \pm 0.014$; the measured $r_{core}$ for Princeton pipe data is

97   $r_{core}^{(\infty)} \approx 0.67 \pm 0.3$, which is believed to be the asymptotic value at high *Re*. From the

98   matching condition, $\kappa = \rho y_{buf}^+ / y_{sub}^{+2}$, we obtain $y_{buf}^{+(\infty)} \approx 44.7$. Fig. 3 compares our

99   predicted MVP with Princeton pipe data using these four parameter values, showing

100  uniform agreement of the entire profile with 99% accuracy for *Re* up to forty million.

101  This unprecedented accuracy, based on objectively measured physical parameters,

102  supports that turbulence in pipe indeed admits an analytic solution!

Note, however, that the core-layer thickness for DNS channel data at moderate $Re$ (Fig.2b) shows notable difference from Princeton pipe data: $r_{core}^{DNS} \approx 0.27$. We attribute it to finite $Re$ effect which can now be studied. Define $r_{core}^{DNS} \approx r_{core}^{(\infty)}(1-\varepsilon_c)$, then departure at DNS $Re$ ($Re_\tau \approx 10^3$) gives $\varepsilon_c \approx 0.6$. Furthermore, we believe that the $Re$-effect in the bulk flow ($r_{core}$) would influence the overlap region ($y_{buf}^+$), leading to a change in $\rho = \rho_0(1+\varepsilon_\rho)$. Fig.1(b) indeed shows evidence of such departure with $\varepsilon_\rho \approx 0.1$, which yields a higher plateau of 0.035. The $Re$-effect is further validated by the measured centreline velocity $U_c^+$ and average velocity $\overline{U^+}$, and skin friction coefficient $C_f$ derived from Eq.(2) (see the Method). Fig.3b, using a compensated plot against asymptotic high $Re$ result, shows clearly that a transition takes place around $Re_\tau^{(crit)} \approx 5000$. Furthermore, a simple model for linear dependence of $\varepsilon_\rho$ and $\varepsilon_c$ on $Re$ below $Re_\tau^{(crit)}$ yields an accurate description of friction coefficients for all pipe data.

The theory also derives several empirical constants of engineering interest. A $Re$-independent quantity for pipe can be predicted: $\left(U_c^+ - \overline{U^+}\right)^{Pipe} \approx 4.3$, which is very close to the empirical value (see Fig.3b), improving Pope's derivation[3], i.e. $1.5/\kappa$ (3.3, using $\kappa \approx 0.45$). Furthermore, we derive a new formula for turbulent pipe for a widely useful relation between $Re_\tau$ and $Re$: our result, $Re_\tau \approx 0.26 Re/\ln Re$, should replace the popular empirical formula[3]: $Re_\tau \approx 0.09 Re^{0.88}$. The latter has an error of up to 25% at $Re_\tau = 5\times 10^5$ according to Princeton data; ours is less than 1%. We also predict a constant difference of centreline velocity between channel and pipe:

$U_c^{+Pipe} - U_c^{+CH} \approx 1.1$, in sharp contrast to [19] which predicts a notable *Re*-dependence.

In summary, we have achieved an accurate description of turbulent mean velocity, which differentiates the effect of geometry (pipe versus channel) with a single integer (m=5 versus m=4), and identified four physical parameters which are *not* fitting parameters, but measured systematically from accurate experimental/numerical data. We obtain also the first theory which predicts finite *Re* effect, and discovers a transition at $\text{Re}_\tau^{(crit)} \approx 5000$ for the friction factor. The most notable outcome is the universality of $\kappa$; it resets the status of the Karman constant to be intimately related to universal small-scale dynamics of fully developed bulk flow turbulence, even possibly to the Kolmogorov constant[24].

The conceptual framework developed here is encompassing and goes far beyond the classical mixing length theory, as the mixing length is used here to reveal the *complete* multi-layer structure which is a general feature of a wide class of wall-bounded turbulent flows. We also show a procedure for extracting quantitative information from empirical data, which can be adapted to a variety of turbulent flows. Indeed, the analysis has been successfully extended to incompressible, compressible and rough-wall turbulent boundary layers, and our results will soon be communicated. Another interesting application underway is turbulent Rayleigh-Benard convection[25]; the mean profiles (of velocity and temperature) can now be quantitatively described by a symmetry study of the mixing length and one or more order functions associated with temperature fluctuations.

## Method

Analytic expressions for mean velocities can be derived from Eq.(2), to yield a systematic procedure for determining four physical model parameters. In the outer region, Eq.(1) yields: $dU^+/dr \approx -\sqrt{r}/\ell_M$. Then, using Eq.(2) for $y^+ \gg y^+_{buf}$:

$$U^{+(Outer)}(r) = U^+(0) - \frac{1}{\kappa} f(r, r_{core}) = U_c^+ - \frac{m}{\kappa} Z_{core} \int_0^r \frac{r' dr'}{(1-r'^m)(r'^2 + r_{core}^2)^{1/4}}, \quad (4)$$

where $f(r, r_{core})$ characterizes the bulk mean flow and depends only on one parameter: $r_{core}$. Once $r_{core}$ is specified, it is a simple least-squares problem to derive optimal values of $\kappa$ and $U_c^+$ from a set of measured mean velocities, $U^{+EXP}(r_i)$. The determination of $r_{core}$ is realized using the relative error function,

$$\sigma_U = \frac{1}{N} \sum \left(1 - U^+(r_i)/U^{+EXP}(r_i)\right)^2,$$

which measures how close the function $f(r, r_{core})$ describes the real data $U^{+EXP}(r_i)$. Note $r_{core}$ and $\kappa$ are determined by independent criteria. The validity of the method is clearly demonstrated in Fig.2.

Analysis of empirical data shows that both $r_{core}$ and $\rho$ have notable $Re$-dependence, while $\kappa$ and $y^+_{sub}$ are remarkably universal. To further elaborate on this, analytic expression for the centreline velocity is derived below. Let us write:

$$U_c^+(Re_\tau) = U_{buf}^+ + \Delta U_{bulk}^+ + \Delta U_{core}^+ \approx \int_0^{y^+_{buf}} \frac{-1+\sqrt{4\ell_M^{+2}+1}}{2\ell_M^{+2}} dy^+ + \left(\int_{y^+_{buf}}^{y^+_{core}} + \int_{y^+_{core}}^{Re_\tau}\right) \frac{\sqrt{r}}{\ell_M^+} dy^+, \quad (5)$$

where the near-wall contribution is expressed by $U_{buf}^+ = U^+(y^+_{buf})$, the bulk by $\Delta U_{bulk}^+$, and the core layer by $\Delta U_{core}^+$, with $y^+_{core} = (1-r_{core})Re_\tau$. Finite $Re$ effect is described by two small parameters, $\varepsilon_\rho$ and $\varepsilon_c$, defined as $\rho = \rho_0(1+\varepsilon_\rho)$ (hence 164    $y_{buf}^+ = y_{buf}^{+(\infty)}(1-\varepsilon_\rho))$ and $r_{core} = r_{core}^{(\infty)}(1-\varepsilon_c)$. A perturbation analysis yields:

165    $U_{buf}^+(\varepsilon_\rho) \approx y_{buf}^{+(\infty)}(0.33-0.20\varepsilon_\rho)$. Substitute $\ell_M^{(bulk)} \approx \kappa(1-r^m)/m$ into (6), to obtain

166    $\Delta U_{bulk}^+ \approx [\ln(y_{core}^+/y_{buf}^+)+\Lambda(r_{core})]/\kappa$, where $\Lambda(r) = \psi(r_{buf})-\psi(r)$ ($r_{buf} = 1 - y_{buf}^+/\text{Re}_\tau \approx 1$

167    for high $Re$) and $\psi(r)$ has an analytic expression [26]. Next, substitute

168    $\ell_M^{(core)} \approx \kappa\left(1+(r_{core}/r)^2\right)^{1/4}/(5Z_{core})$ into (6), to obtain $\Delta U_{core}^+ \approx 5Z_{core} r_{core}^{3/2}/(2\kappa)$. Summarizing

169    all the terms above, we obtain a first-order approximation for the centreline velocity:

170    $$U_c^{+Pipe(CH)} \approx U_c^{+Pipe(CH)(\infty)} - 6.72\varepsilon_\rho + 1.51(1.68)\varepsilon_c \qquad (6)$$

171    where $U_c^{+Pipe(CH)(\infty)} \approx \ln(\text{Re}_\tau)/\kappa + 8.37(7.28)$. This yields, for the first time, the centreline

172    velocity difference between a pipe and a channel at the same $Re$:

173    $U_c^{+Pipe} - U_c^{+CH} \approx 1.09 - 0.17\varepsilon_c$. Finally, the average velocities also have an analytic

174    expression:

175    $$\overline{U^+}^{Pipe(CH)} \approx \overline{U^+}^{Pipe(CH)(\infty)} - 6.84\varepsilon_\rho + 0.26(0.36)\varepsilon_c. \qquad (7)$$

176    where $\overline{U^+}^{Pipe(CH)(\infty)} \approx \ln(\text{Re}_\tau)/\kappa + 4.1(5.0)$. This allows an evaluation of the bulk velocity

177    Reynolds number in pipe: $\text{Re} = 2\overline{U}/\nu = 2\overline{U^+}\text{Re}_\tau$, and friction coefficient:

178    $$C_f^{Pipe} = 8\left(\overline{U^+}^{Pipe}\right)^{-2} \approx C_f^{Pipe(\infty)}[1+(13.68\varepsilon_\rho - 0.52\varepsilon_c)\sqrt{C_f^{Pipe(\infty)}/8}]. \qquad (8)$$

179    where $C_f^{Pipe(\infty)} = 8\left(\overline{U^+}^{Pipe(\infty)}\right)^{-2}$. A simple linear model for $\varepsilon_{\rho(c)}$ is introduced as:

180    $$\varepsilon_{\rho(c)} = \varepsilon_{\rho(c)}^{(0)}(\text{Re}_\tau^{(crit)} - \text{Re}_\tau)/(\text{Re}_\tau^{(crit)} - 1000), \qquad (9)$$

181    where $\varepsilon_\rho^{(0)} \approx 0.1$ and $\varepsilon_c^{(0)} \approx 0.6$ are derived from empirical data.

182

## References

1. Prandtl, On fluid motions with very small friction (in German). *Third International Mathematical Congress*. Heidelberg, 484–491 (1904).

2. von Karman, T. Mechanische Ahnlichkeit und Turbulenz. In *Proc. Third Int. Congr. Applied Mechanics*, Stockholm. 85-105 (1930).

3. Pope, S.B. *Turbulent flows*. 264-295 (Cambridge University Press, 2000).

4. Smits A. J., McKeon B. J., Marusic I., High Reynolds number wall turbulence. *Annu. Rev. Fluid Mech.* 43:353-75 (2011).

5. Nagib, H.M. and Chauhan, K.A. Variations of von Kármán coefficient in canonical flows. *Phys. Fluids.* **20**, 101518 (2008).

6. Marusic, I., *et al.* Wall-bounded turbulent flows at high Reynolds numbers: Recent advances and key issues. *Phys. Fluids.* **22**, 065103 (2010).

7. She, Z.S., Chen, X., Wu, Y. and Hussain, F. New perspectives in statistical modeling of wall-bounded turbulence. *Acta Mechanica Sinica.* **26**, 847-861 (2010); She, Z.S., Chen, X. and Hussain, F. A Lie-group derivation of a multi-layer mixing length formula for turbulent channel and pipe flows, arXiv:1112.6312 (2011).

8. Wilcox, D.C. *Turbulence Modeling for CFD*. (DCW Industries, 2006), 181-182.

9. Reynolds, O. An experimental investigation of the circumstances which determine whether the motion of water shall be direct or sinuous, and the law of resistance in parallel channels. *Phil. Trans. R. Soc. London, Ser. A.* **174**, 935-982 (1883).

10. Zagarola, M.V. and Smits, A.J. Scaling of the mean velocity profile for turbulent pipe flow. *Phys. Rev. Lett.* **78**, 239-242 (1997).

11. Swanson, C., Julian B., Ihas, G.G. and Donnelly, R., Pipe flow measurements over a wide range of Reynolds numbers using liquid helium and various gases, *J. Fluid Mech.* 461, 51-60 (2002).

12. Zanoun, E.S., Durst, F. and Nagib, H. Evaluating the law of the wall in two-dimensional fully developed turbulent channel flows. *Phys. Fluids.* **15**, 3079-3089 (2003).

13. B. J. McKeon, J. Li, W. Jiang, J. F. Morrison, and A. J. Smits. *J. Fluid Mech.* **501**, 135 (2004); the data are available at http://gasdyn.princeton.edu/data/e248/mckeon_data.html.

14. Talamelli, A. *et al.* CICLoPE—a response to the need for high Reynolds number experiments. *Fluid Dyn. Res.* **41**, 021407 (2009).

15. Barenblatt, G.I. Scaling laws for fully developed turbulent shear flows. Part 1. Basic hypotheses and analysis. *J. Fluid Mech.* **248**, 521-529 (1993).

16. Barenblatt, G.I., Chorin, A.J. and Prostokishin V.M. Scaling laws for fully developed turbulent flow in pipes: Discussion of experimental data. Proc. Natl. Acad. Sci. 94, 773-776 (1997).

17. Zagarola, M.V., Perry, A.E. and Smits, A.J. Log laws or power laws: The scaling in the overlap region. *Phys. Fluids.* **9**, 2094-2100 (1997).

18. George, W.K. Is there a universal log-law for turbulent wall-bounded flow? *Phil. Trans. R. Soc. London, Ser. A.* **365**, 789-806 (2007).

19. L'vov, V.S., Procaccia, I. and Rudenko, O. Universal model of finite Reynolds number turbulent flow in channels and pipes. *Phys. Rev. Lett.* **100**, 050504 (2008).

20. Kadanoff, L. P. More is the same; phase transitions and mean field theories. *J. Stat. Phys.* **137**, 777–797 (2009).

229    21. Iwamoto, K., Suzuki, Y. and Kasagi, N. Database of fully developed channel flow.
230    *THTLAB Internal Report*. No. ILR-0201 (2002).

231    22. Hoyas, S. and Jimenez, J. Scaling of the velocity fluctuations in turbulent channels
232    up to Retau = 2003. *Phys. Fluids.* **18**, 011702 (2006).

233    23. Wu, X.H. and Moin, P. A direct numerical simulation study on the mean velocity
234    characteristics in turbulent pipe flow. *J. Fluid Mech.* **608**, 81-112 (2008).

235    24. Lo, T.S., L'vov, V.S., Pomyalov A. and Procaccia, I. Estimating von-Karman's
236    constant from homogeneous turbulence. *Europhys. Lett.* **72**, 943 (2005).

237    25. Ahlers, G., Grossmann, S. and Lohse, D. Heat transfer and large scale dynamics in
238    turbulent Rayleigh-Bénard convection. *Rev. Mod. Phys.* **81**, 503–537 (2009).

239    26. For pipe flow (m=5), the explicit form is

240    $\psi^{Pipe}(r) = 2\ln(1+\sqrt{r}) + \frac{\sqrt{5}-1}{4}\ln\varphi^+(r) + \frac{\sqrt{5}+1}{4}\ln\varphi^-(r) + \frac{2\sqrt{5}}{c^-}\arctan\frac{c^-\sqrt{r}}{A^-(r)} - \frac{2\sqrt{5}}{c^+}\arctan\frac{c^+\sqrt{r}}{A^-(r)}$, where

241    $\varphi^+(r) = \frac{A^+(r)-B^+(r)}{A^+(r)+B^+(r)}$; $\varphi^-(r) = \frac{A^+(r)-B^-(r)}{A^+(r)+B^-(r)}$, $A^+(r) = 2(1+r); A^-(r) = 2(1-r);$

242    $B^+(r) = (\sqrt{5}+1)r; B^-(r) = (\sqrt{5}-1)r;$  $c^+ = \sqrt{10+2\sqrt{5}}; c^- = \sqrt{10-2\sqrt{5}}.$

243

## End Notes

245

**Acknowledgements** We thank G. Ahlers, E. Bodenschatch, N. Goldenfeld, H. Swinney for helpful discussions. This work was supported by the National Natural Science Foundation of China (90716008 and 10921202), MOST 973 Project (2009CB724100). Part of the work was completed during our visit at KITP of UCSB during "Turbulence program", supported by the National Science Foundation of US under Grant No. NSF PHY05-51164, which is gratefully acknowledged.

**Author Contribution**

Z.S.S. established the theoretical framework, and directed the research in collaboration with F.H.. X.C. and Z.S.S. developed the analysis, and X.C. measured the physical constants. Y.W. performed quantitative study of the model and compared the predictions with the empirical data. Z.S.S., X.C. and F.H. wrote the manuscript. All authors extensively discussed the results and interpretations, and commented on the manuscript.

**Author Information** The authors declare no competing financial interests. Correspondence and requests for materials should be addressed to Z.S.S. (she@pku.edu.cn).

**Figure Legends**

**Figure 1.** Validation of the multi-layer description of the mixing length profile and the determination of the near-wall constants. Three sets of best resolved DNS data are used: two channel flows from Iwamoto et al.[21] at $Re_\tau \approx 650$ (black circles) and Hoyas and Jimenez[22] at $Re_\tau \approx 940$ (black squares), and one pipe flow of Wu and Moin[23] at $Re_\tau \approx 1142$ (blue triangles). (a): Compensated plot of the mixing length, $\ell_M^{DNS}/(1-r^m)$, illustrates four layers: sublayer, buffer, log-bulk and core. Two plateaus ($m=4$ for channel and $m=5$ for pipe) illustrate the bulk flow structure, with the same $\kappa = 0.45$. The solid lines are predictions of Eq.(1) using the measured $\kappa$ and other parameters. (b): Near-wall behaviour compared to theoretical prediction, $\rho_0/y_{sub}^{+3/2} \approx 0.0315$ (dashed line), finite-$Re$ description of composite solution (solid line), and corresponding buffer-layer description (dash-dotted line), which measure two constants. The functions[7] $B^{s-b} = (1+(y^+/y_{sub}^+)^4)^{1/8}$ and $B^{b-l} = (1+(y^+/y_{buf}^+)^4)^{-1/4}$.

**Figure 2.** Determination of the Karman constant, $\kappa$, and the core-layer thickness, $r_{core}$. (a) Plot of measured $U_d^+(r) = U^+(0) - U^{+EXP}(r)$ versus theoretical function $f(r, \bar{r}_{core})$, to illustrate a good linear relationship consistent with the measured universal constant $\kappa \approx 0.45$, for both channel and pipe flows for a wide range of $Re$. Inset shows the measurement for all Princeton data for $Re_\tau > 5000$. Star is the final measured value: $\bar{r}_{core}^{EXP} \approx 0.67$ for Princeton data at relatively high $Re$, while $\bar{r}_{core}^{DNS} \approx 0.27$ for DNS data at moderate $Re$, and $\kappa \approx 0.45$ in both cases. (b) Result of applying the procedure to

287  four sets of data: one theoretical profile generated from Eq.(1) and Eq.(2) to test the
288  validity of the procedure, second profile with noise added to the former to test the
289  robustness of the method, third profile from DNS channel simulation, and Princeton
290  pipe data[13]. Note that the DNS profile follows closely the same error variation pattern as
291  the theoretical profile, validating the theory with a clear determination of $r_{core}$ and
292  yielding a reliable estimate of $\kappa$ and $U_c^+$. On the other hand, the profile with small
293  random noise (0.5% ) yields an error variation pattern, closely resembling Princeton
294  pipe data, which yields a scatter (20-30%) of the estimated $r_{core}$, but a much smaller
295  uncertainty of $\kappa$ (only 3%). This explains the scatter in the measured $r_{core}$ presented
296  in the inset of (a), and the slight variation of $\kappa$.

297

298  **Figure 3.** (a): Theoretical (solid lines) and measured MVPs, which are staggered
299  vertically by five units for clarity. Inset shows the relative errors,
300  $(U^{EXP}/U^{Theory} - 1) \times 100$ %. Note that errors in our theoretical predictions (red solid
301  symbols) are uniformly within 1%. Also included is the recent model of L'vov,
302  Proccacia and Rudenko (LPR[19]) (blue open symbols) - the only other theoretical model
303  for entire MVP at finite $Re$. The LPR model has three adjustable parameters, the most
304  important choices being $\kappa = 0.415$ for channel and 0.405 for pipe. It reveals
305  systematic deviations at high $Re$. (b): Compensated plot of friction coefficient
306  $C_f^{EXP}/C_f^{Pipe(\infty)}$ shows a transition around $Re_\tau^{(crit)} = 5000$ (black dot line). The dash line
307  indicates $C_f^{EXP} = C_f^{Pipe(\infty)}$, and the dash-dot line is a linear finite $Re$ model for both $\varepsilon_\rho$
308  and $\varepsilon_c$ given by Eq.(8) and (9). Inset compares the present prediction of 309 $\left(U_c^+ - \overline{U^+}\right)^{Pipe} \approx 4.3$ (black dash line) to Princeton data (symbols), far improving previous

310 description in [3] (blue solid line).

**Figure 1.**

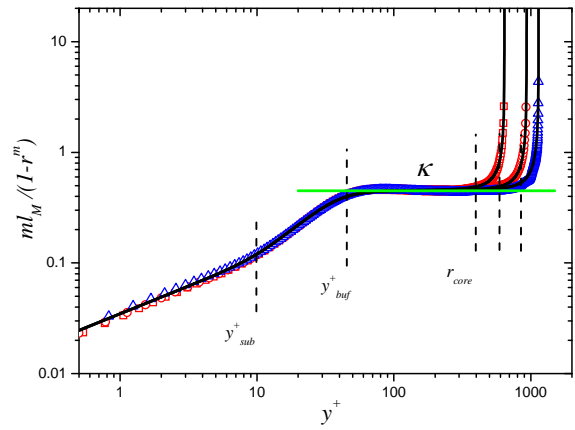

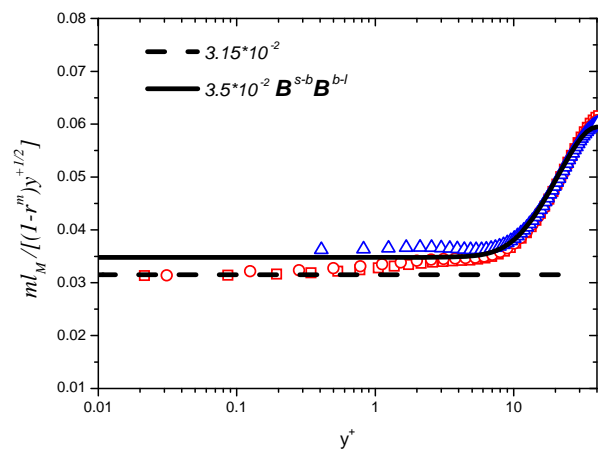

**Figure 2.**

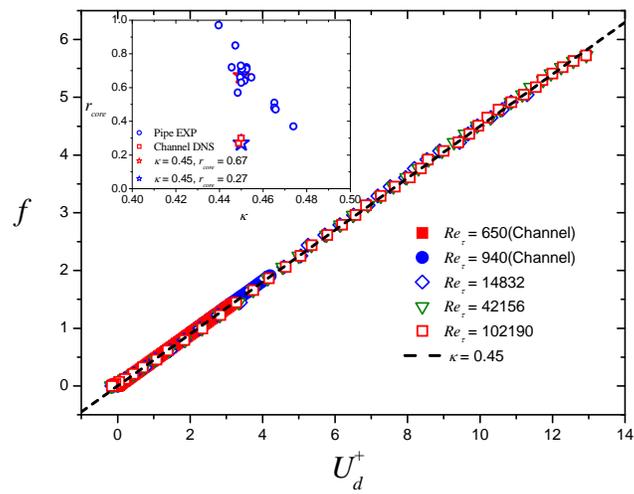

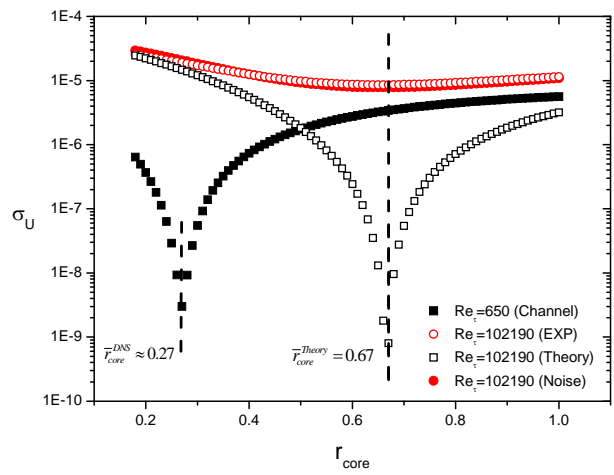

320  **Figure 3.**

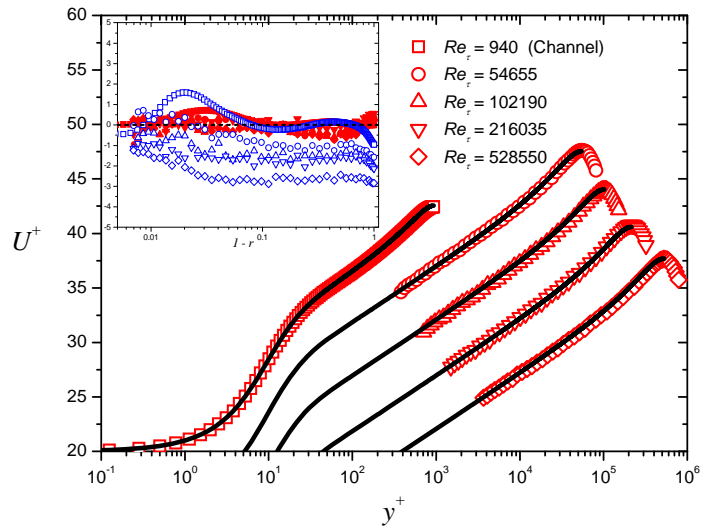

321

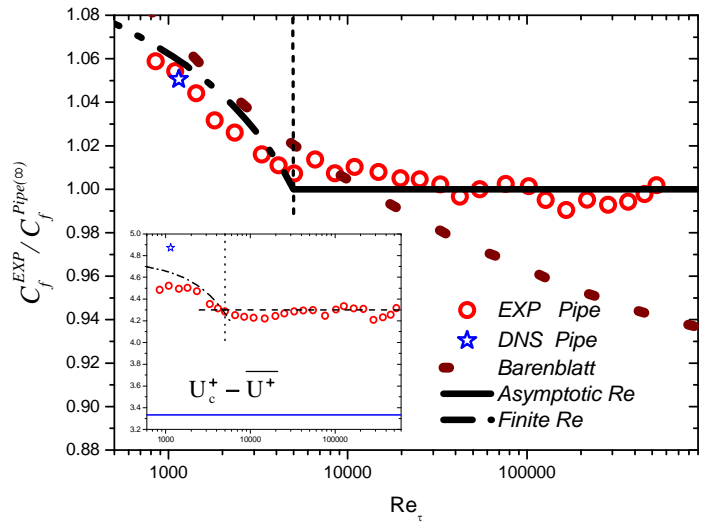

322

323

324

325  **End**